# Impact of invasive metal probes on Hall measurements in semiconductor nanostructures

Jan G. Gluschke,[a] Jakob Seidl,[a] H. Hoe Tan,[b] Chennupati Jagadish,[b] Philippe Caroff [b,c] and Adam P. Micolich *[a]

Recent advances in bottom-up growth are giving rise to a range of new two-dimensional nanostructures. Hall effect measurements play an important role in their electrical characterization. However, size constraints can lead to device geometries that deviate significantly from the ideal of elongated Hall bars with currentless contacts. Many devices using these new materials have a low aspect ratio and feature metal probes that overlap with the semiconductor channel. This can lead to a significant distortion of the current flow. We present experimental data from InAs 2D nanofin devices with different Hall probe geometries to study the influence of Hall probe length and width. We use finite-element simulations to further understand the implications of these aspects and expand the scope to contact resistance and sample aspect ratios. Our key finding is that invasive probes lead to a significant underestimation in the measured Hall voltage, typically of the order of 40-80%. This in turn leads to a subsequent proportional overestimation of carrier concentration and an underestimation of mobility.

## 1. Introduction

The Hall effect[1] has a long history, both as a source of novel states of matter, e.g., the integer,[2] fractional,[3] and spin[4] Hall effects, and as an essential characterization tool for semiconducting materials.[5] Hall studies are particularly amenable to the recent explosion in 2D nanomaterials that began with graphene,[6,7] extended into transition metal dichalcogenides,[8,9] e.g., MoS$_2$,[10,11] MoSe$_2$,[12] WS$_2$,[10] and WSe$_2$,[13] and has most recently moved to III-V semiconductors[14–24] aiming to meet demand for more complex III-V nanostructure geometries in fields such as quantum computing.[25–29]

We have recently worked on one of the new 2D morphologies of III-V semiconductor emerging from advances in bottom-up growth approaches. Our selective-area epitaxy grown InAs 2D nanofins are rectangular – typically a few microns long and wide, and 50-100 nm thick.[14] Electrical contacts for device characterization are fabricated with electron-beam lithography (EBL) and subsequent metal deposition. Hall probes made in this way overlap with the nanostructure and are typically a few hundred nanometres wide. This is significant compared with the size of the nanostructure. Figure 1a shows a typical device.

This Hall effect geometry is far from the ideal of a high aspect ratio, rectangular 'Hall bar' with recessed, non-invasive, currentless Hall voltage probes that are well separated from the source and drain contacts.[5] Instead, in 2D nanomaterial Hall devices, the aspect ratio is low because it is determined by the intrinsic nanostructure geometry and often closer to 1:1. Furthermore, the metal Hall probes overlapping the conduction channel form a parallel current path through the nanostructure segment, which perturbs the current density. These non-idealities influence the result of a Hall voltage measurement but are generally ignored in characterizations of new 2D nanomaterials. This can lead to a significant misestimate of the Hall voltage, and subsequently, the extracted material parameters such as carrier density and mobility. It is thus important to systematically study and quantify the impact of non-idealities in 2D nanostructure Hall devices to correct for their impact and enable more accurate material characterization.

Here, we systematically investigate the impact of sample and contact geometry for 2D nanomaterials with invasive metal contacts on classical Hall effect measurements. We used InAs nanofin Hall devices with multiple Hall probe pairs to measure the effect of contact geometry on a single nanofin. We obtain up to 2.5-fold differences in measured Hall voltage for different contact geometries on the same device. We extend our experimental study using finite-element modelling to better understand and quantify the impact of sample aspect ratio, contact resistance and the width and length of the Hall probes. Based on our data, we provide recommendations on how to design devices to reduce current perturbation and how to estimate a Hall voltage correction factor for non-ideal devices. We estimate that direct Hall voltage measurements on typical

a. School of Physics, University of New South Wales, Sydney NSW 2052, Australia
b. Department of Electronic Materials Engineering, Research School of Physics, The Australian National University, Canberra, ACT 2601, Australia
c. Microsoft Quantum Lab Delft, Delft University of Technology, 2600 GA Delft, The Netherlands

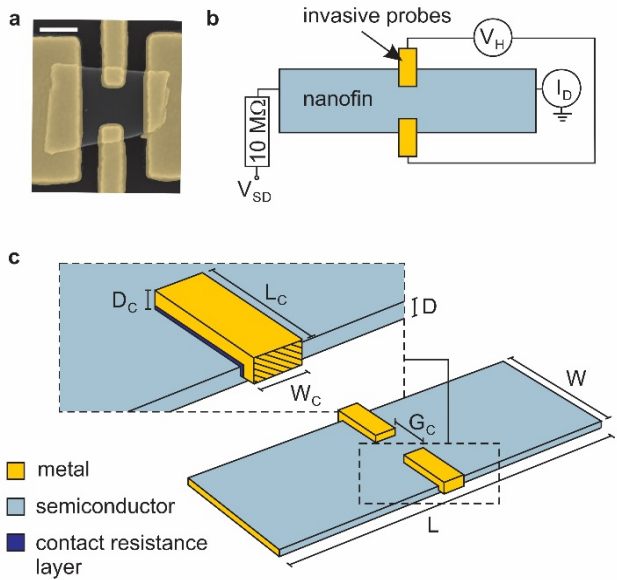

**Figure 1.** (a) Scanning electron microscopy (SEM) image of a typical nanofin device with two metal Hall probes (scale bar = 500 nm). (b) Schematic of the electrical measurement circuit. (c) Schematic of a model approximating a nanofin device in our simulations defining the width $W$, length $L$ and height $H$ for the 2D structure as well as the width $W_C$, length $L_C$, and height $H_C$ for the probes. The gap between the Hall probes is $G_C$. Note that $W_C$ runs perpendicular to $W$ and $L_C$ runs perpendicular to $L$ for convenience of discussion.

2D nanostructure devices with invasive contacts underestimate the real Hall voltage by around 40-80%. This leads to a proportional overestimation of carrier concentration and underestimation of the carrier mobility.

## 2. Methods

### 2.1 Nanofin growth and device processing

InAs nanofins with a high aspect ratio were chosen to isolate the impact of invasive probes on the Hall measurements from effects arising from the nanofin geometry. Full details are given in Ref. 14, but briefly, the nanofins are 1.05 μm wide, and approximately 5 μm long, 70-110 nm thick and grown on an InP(111)B substrate by metal-organic vapor phase epitaxy. The growth was templated by pre-patterned trenches defined by electron beam lithography and dry-etching in a $SiO_x$ mask layer on the substrate.[14] Individual nanofins were carefully mechanically transferred to a Si device substrate with a 100 nm $SiO_2$ layer and prefabricated alignment markers. Electrical contacts were patterned using EBL. The contact regions were exposed to a 30 s oxygen plasma etch (350 mTorr, 50 W) to remove resist residue, and then $(NH_4)_2S_x$ passivation solution at 40°C for 120 s immediately prior to contact metal deposition to remove native oxide and improve ohmic contact formation.[30] The contact metal (Ni/Au 5/135 nm) was deposited by thermal evaporation. Narrow probe leads can suffer discontinuities during lift-off reducing their yield. The devices were inspected under a scanning electron microscope (SEM) prior to electrical characterization to confirm contact alignment. Care was taken to limit the electron-beam exposure during SEM to avoid significant changes in electrical characteristics.[31,32]

### 2.2 Electrical characterization

Figure 1b shows a schematic of the measurement configuration. Hall voltage $V_H$ data was obtained using SR830 lock-in amplifiers with a current $I_D$ at 77 Hz between source and drain passed via a 10 MΩ series resistor to maintain a constant current of 100 nA (typical sample resistance ~5 kΩ). Measurements were performed with the device at a temperature of 20 K to prevent strong quantum conductance fluctuations[14] from obscuring the Hall data. We used an Oxford Instruments Heliox VL ³He system with a 2 T magnet in a Wessington CH-120 helium dewar to achieve this.

### 2.3 Finite-element modelling

The simulations were performed using COMSOL Multiphysics 5.1 with the electric currents (ec) module in a stationary study. The nanofin was modelled as a cuboid with length $L = 5$ μm, width $W = 1$ μm and thickness $D = 75$ nm unless otherwise specified. The structure was assigned the anisotropic conductivity tensor:[5]

$$\sigma = \mu n e \begin{pmatrix} \dfrac{1}{1+\mu^2 B^2} & -\dfrac{\mu B}{1+\mu^2 B^2} & 0 \\ \dfrac{\mu B}{1+\mu^2 B^2} & \dfrac{1}{1+\mu^2 B^2} & 0 \\ 0 & 0 & 1 \end{pmatrix} \quad (1)$$

with carrier mobility $\mu$, carrier concentration $n$, electron charge $e$ and magnetic field $B$.[33] We chose $\mu = 3100$ cm²/(V·s) and $n = 2 \times 10^{17}$ cm⁻³ as default values based on our earlier study of InAs 2D nanofins.[14] However, the effects leading to a reduced measured Hall voltage discussed in this paper only depend device geometry and the ratio of sample to Hall probe conductivity and not on the specific values of $n$ and $\mu$. We expect our model to hold for all devices where the Drude approximation, i.e., Equation 1, is valid.

Figure 1c shows a schematic of a modelled device with one pair of metallic Hall probes with contact length $L_C$ and width $W_C$. The gap between the two probes is $G_C = W - 2L_C$. The contact height $D_C$ is fixed at 100 nm. The contacts were assigned an isotropic conductivity of $\sigma_C = 8.5 \times 10^8$ S/m to approximate commonly used Au contacts.[34] This is approximately 5 orders of magnitude higher than the nanofin conductivity. The finite thickness of the nanofins inevitably means contacts on the top surface need to step down the edges onto the substrate to continue to the external circuitry. The Hall probes in the simulation extend half-way down the side of the nanofin to account for this. Regardless, we find that the extent of side facet coverage does not have a significant impact on the modelled Hall voltage (see Supporting Information Fig. S1). Note that the Hall voltage extracted from the simulation depends on the relative rather than absolute dimensions of the simulated device. This means that the considerations presented for e.g. 5 μm × 1 μm equally apply for 50 nm × 10 nm devices provided that all other dimensions (e.g. Hall probes) are scaled accordingly.

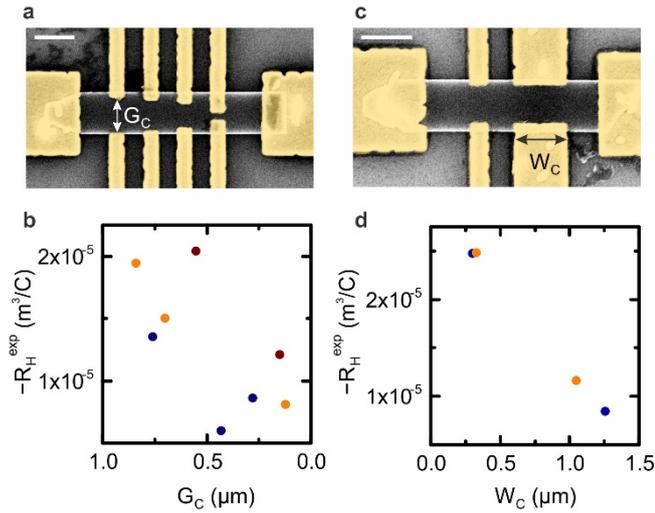

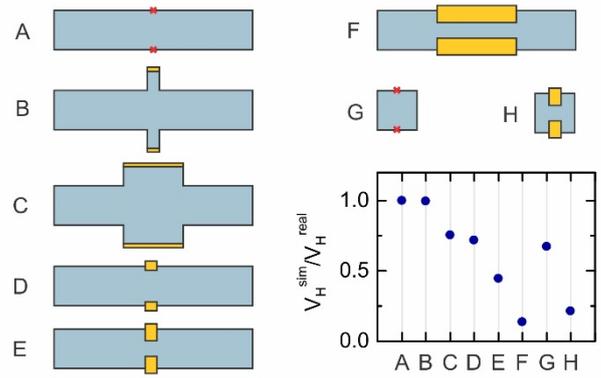

**Figure 3.** Simulation of the Hall voltage $V_H^{sim}$ for different device and contact geometries: high aspect ratio nanofins with 'ideal' contactless probes (A), a classic Hall bar (B), wide recessed Hall probes (C), device with short metal probes (D), long metal probes (E), wide metal probes (F), low aspect-ratio nanofin with 'ideal' contactless probes (G), and invasive contacts (H). $V_H^{sim}$ is normalized to the 'real' Hall voltage $V_H^{real}$.

**Figure 2.** (a) SEM image of a nanofin device with four independent pairs of Hall probes for testing effect of probe gap $G_C$ (scale bar = 1 µm). (b) Measured Hall coefficient $R_H^{exp}$ vs $G_C$ for three different devices represented by different colours (device from (a) shown in orange). (c) SEM image of a nanofin device with two independent pairs of Hall probes for testing the effect of contact width $W_C$ (scale bar = 1 µm). (d) $R_H^{exp}$ vs $W_C$ for two different devices (device from (c) shown in orange). The data point width is equal to the uncertainty in $G_C$ and $W_C$ respectively.

A 10 nm layer with isotropic conductivity $\sigma_{CR} \leq \sigma_C$ has been placed between the metal contact and the semiconductor structure to model contact resistance (blue in Figure 1c). However, we consider no contact resistance $\sigma_{CR} = \sigma_C$ as our default case unless specified otherwise. The two surfaces at the ends of the nanofin were modelled as electrical terminals with the source supplying a constant current $I_D = 100$ nA and the drain set at 0 V. The outer side of each Hall contact was set as a floating potential to extract the Hall voltage (shaded area in Figure 1c). We used $B = 0.1$ T unless otherwise indicated and simulations were computed to a relative tolerance of $10^{-5}$.

### 2.4 Clarification of terminology

In this paper we discuss three different Hall voltages. The 'real' Hall voltage $V_H^{real}$ is calculated analytically from the input parameters $I_D$, $B$, $n$, and $D$:[5]

$$V_H^{real} = \frac{I_D B}{nDe} \quad (2)$$

This is the Hall voltage value that, by definition, yields the correct carrier concentration. We use it to normalise our other two Hall voltages $V_H^{sim}$ and $V_H^{exp}$ for clearer analysis. The second is the Hall voltage $V_H^{sim}$ extracted from the simulations. The model simulates a measurement where $V_H^{sim}$ is the difference in electrical potential of two Hall probes. Some models in Section 3.2 and 3.3 do not have Hall probes. Here, $V_H^{sim}$ corresponds to the potential difference between two points at the modelled sample's edge. The third Hall voltage is the measured Hall voltage $V_H^{exp}$, which is directly taken from experimental measurements performed on real samples. If a measurement or simulation is 'ideal' then we obtain $V_H^{sim}/V_H^{real} = V_H^{exp}/V_H^{real} = 1$. Deviations from 1 indicate non-ideality. $R_H^{sim}$, $R_H^{exp}$ and $R_H^{real}$ are the corresponding Hall coefficients.

## 3. Results and discussion

### 3.1 Hall effect measurements in practice

Sample and Hall-probe geometry can both vary widely in the characterization of 2D nanostructures. The high aspect ratio of our selective-area epitaxy grown nanofins (5 × 1 µm²) allows the placement of multiple Hall probes with differing probe gap $G_C$ and width $W_C$. This enables us to measure $V_H^{exp}$ for different probe geometries without the effect of sample-to-sample variations. Figure 2a shows a device with four Hall probes. All contacts are ~350 nm wide but the gap between the probes $G_C$ varies between 120 and 840 nm. For an ideal Hall device all probes should measure the same Hall voltage. Figure 2b shows that this is not the case. The measured Hall coefficient $-R_H^{exp} = -V_H^{exp}D/(I_D B)$ is significantly smaller for the probe pairs with smaller gaps across all three devices (different colour data points in Fig. 2b).

Figure 2c shows a device with probe pairs of different width $W_C$. The measured Hall coefficients for different $W_C$ are shown in Figure 2d. $R_H^{exp}$ is almost three times larger for the narrow probe pair in both measured devices despite the contact gap $G_C$ being the same. This would yield a nearly three-fold difference in extracted carrier concentrations for the same sample. The measurements demonstrate that probe geometry has a significant impact on the outcome of a Hall measurement. In the following sections we use finite-element modelling to investigate the impact of contact and sample geometry on probed Hall voltage.

### 3.2 Sample geometry

Figure 3 gives an overview of how different sample and contact geometries influence the simulated probed Hall voltage $V_H^{sim}$. All extracted Hall voltages were normalized to $V_H^{real}$, such that an ideal modelled Hall device would return $V_H^{sim}/V_H^{real} = 1$. Geometry A is a 'contactless' $5 \times 1$ µm$^2$ rectangular sample with $V_H^{sim}$ extracted directly from the electrical potential difference between the two points at the centre of the side surfaces. We obtain $V_H^{sim} = V_H^{real}$ here, as expected. We get the same result for the traditional Hall bar geometry with currentless recessed contacts (Geometry B). In contrast, Geometries D-F feature invasive metal contacts overlapping the semiconductor channel. Here, the simulations show that $V_H^{sim}$ is significantly smaller than $V_H^{real}$. The reduction is particularly pronounced for the wide contacts in Geometry F where $V_H^{sim}$ is less than 15% of $V_H^{real}$. This would lead to an underestimation of carrier concentration by a factor of > 6 in a real measurement. Note that recessed probes also become invasive if the probes are sufficiently short and wide relative to the sample dimensions (Geometry C). Such geometries are relevant to, e.g., nanocross devices[27,35] and are discussed in Section 3.5 and in detail in the Supporting Information S2-S3.

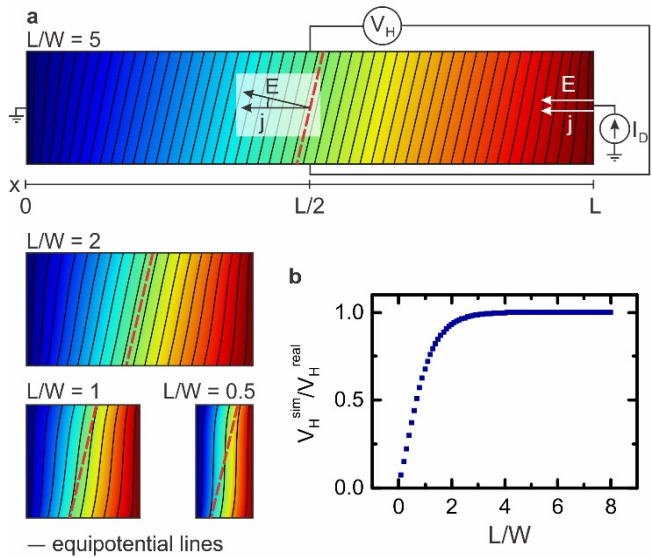

**Figure 4.** (a) Equipotential lines for four different aspect-ratio samples $L/W$ at magnetic field $B = 0.75$ T. The red dashed line indicates an equipotential line corresponding to the Hall angle of $V_H^{real}$. (b) transverse voltage $V_H^{sim}$ drop at the sample's centre normalized to $V_H^{real}$ vs sample aspect ratio $L/W$.

In Geometries G and H we consider the effect of reducing the aspect ratio $L{:}W$, which further lowers $V_H^{sim}$. We find a 30% reduction in Hall voltage even without invasive probes (G) for a sample with 1:1 aspect ratio. In the following subsections we will examine the factors causing the reduction in Hall voltage more closely.

### 3.3 Sample aspect ratio

Probing of the Hall voltage too close to the source and drain contacts leads to an underestimate of the Hall voltage.[5,36] This is why the Hall probes are separated from the source and drain probes by at least four times the width of the sample in ideal Hall bar devices.[5,36] Good geometric control is often not achieved in many 2D nanostructures. This leads to measurements with low aspect ratio and poor probe separations being used. A quantitative estimate of the reduction in Hall voltage due to a reduced sample aspect ratio is thus necessary to correct these measurements.

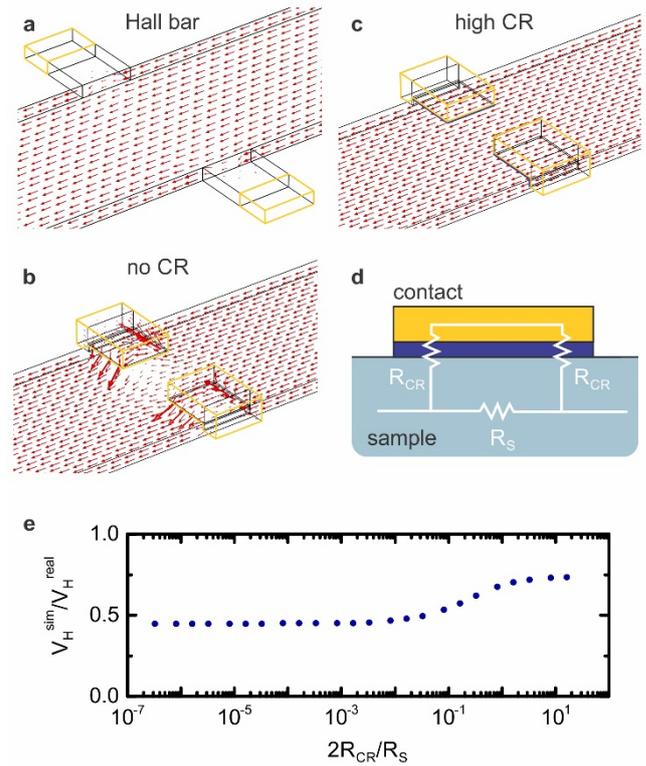

**Figure 5.** Simulation of current density (red arrows) in (a) a classic Hall bar, (b) sample with 'invasive' low contact resistance metal probes, and (c) sample with high contact resistance metal probes. (d) Schematic of the two competing current paths through the sample and the contact in the probed segment. $R_{CR}$ is the contact resistance and $R_S$ is the resistance of the segment below the contact. (e) $V_H^{sim}/V_H^{real}$ for different contact $2R_{CR}/R_S$ ratios.

Figure 4a shows electrical equipotential lines for samples with different aspect ratios $L{:}W$ in a simulated Hall measurement at magnetic field $B = 0.75$ T. The source and drain contacts force the interfaces at both ends to be at fixed potential (0 V and $V_{SD}$ respectively). Here the electric field $\boldsymbol{E}$ is parallel to the current density $\boldsymbol{j}$ and both are perpendicular to the contact interface. The equipotential lines gradually turn as one moves closer to the sample centre until $\boldsymbol{E}$ is at the Hall angle $\Theta = \tan^{-1}(\mu B)$ relative to $\boldsymbol{j}$ at the middle of the Hall bar. The Hall voltage at any position $x$ along the sample is the transverse potential difference between the two edges at that $x$. The red dashed line indicates an equipotential line corresponding to the Hall angle of $V_H^{real}$. If $V_H^{sim}$ is probed too close to the source and drain contacts then $V_H^{sim}$ is significantly smaller than $V_H^{real}$. This matters for low aspect ratio samples because the length is insufficient for $\boldsymbol{E}$ to align with the 'real' Hall angle at the position $x$ where the contacts are located. In Figure 4b we plot the ratio $V_H^{sim}/V_H^{real}$ using the potential difference between opposite edges at $x = L/2$ to identify the aspect ratio where non-ideality becomes significant. We find that the expected Hall voltage $V_H^{sim}$ is within 1% of $V_H^{real}$ for aspect ratio $L{:}W$ above 3.3:1 and within 10% for aspect ratio above 1.8:1. The reduction in $V_H^{sim}$ is more than 30% for aspect ratios below 1:1, i.e., samples with $L < W$. The effect can be further compounded by the effect of metal probes as we see in §3.4. The simulations in the following sections will focus on samples with high aspect ratio (5:1) to isolate the effects of invasive probes from those of sample geometry.

### 3.4 Invasive metal contacts

A homogenous current density is assumed when evaluating a Hall measurement. Recessed, currentless Hall probes are used in traditional Hall bars to ensure that the voltage probes do not interfere with the sample current. Figure 5a shows the modelled current density (red arrows) through a traditional Hall bar confirming there is no significant current perturbation. This is not the case for 'invasive' metal probes that overlap the conduction channel and are used in characterization of many 2D nanostructures; see Figure 5b, where the metal contacts are traced in yellow for clarity. The simulation shows a clear reduction in current density between and beneath the metal contacts. This occurs because the contacts themselves provide the lowest resistance path through this sample segment. The resulting perturbation of the current density significantly reduces $V_H^{sim}$.

In our simulation we use the 10 nm thin interface layer between the nanofin and the metal contact to simulate contact resistance. This enables us to vary the resistance of the current path through the contact. A simplified model of the two current paths through the probed nanofin segment is shown in Figure 5d. We estimate the contact resistance $R_{CR}$ as the resistance of the 10 nm layer with conductivity $\sigma_{CR}$ and area $A_{CR}$ so that $R_{CR} = 10\,\text{nm}/(\sigma_{CR}\cdot A_{CR})$. We compare this to the resistance $R_S = W_C/(\mu ne\cdot L_C\cdot D)$ of the sample segment immediately below the contact. Figure 5e shows $V_H^{sim}/V_H^{real}$ versus $2R_{CR}/R_S$, providing a rough estimate of the ratio of the resistances of the two current paths. The reduction of the Hall voltage is largest where $R_{CR} \ll R_S$. This means that the current path through the contact is the path of lowest resistance (see Figure 5b). The Hall voltage increases as $2R_{CR}/R_S$ approaches 1 and saturates for $R_{CR} \gg R_S$, where current no longer flows through the Hall probes. This case is shown in Figure 5c. In this regime, the potential sensed by each voltage probe is the average electric potential at the nanofin-probe interface. The probed Hall voltage is then the potential difference between the two Hall probes. $V_H^{real}$ can therefore be estimated using:

$$V_H^{real} \approx V_H^{exp/sim}/(0.5 + \frac{G_C}{2W}) \quad (3)$$

We find that the measured Hall voltage is approximately 50% of $V_H^{real}$ for low to moderate contact resistances and just over 70% for high contact resistances for this particular geometry ($W_C = 300\,\text{nm}$, $G_C = 400\,\text{nm}$). In other words, high contact resistance probes make the Hall voltage measurement more accurate and easier to interpret because $V_H^{real}$ can be estimated using Equation 3. The contact resistances are generally very low in a material with a prominent surface accumulation layer like InAs [37]. Thus the resulting current distortion effects need to be taken into account. The impact of the current distortion effects on the Hall measurement is not straightforward and depends on multiple parameters – most significantly on probe width and length as we will discuss in Section 3.6.

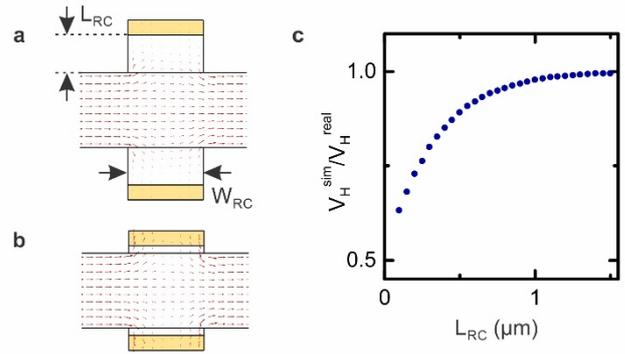

**Figure 6.** Simulated current density for a 5 μm × 1 μm nanofin with 1 μm wide and (a) 0.5 μm and (b) 0.1 μm long recessed contacts. (c) $V_H^{sim}/V_H^{real}$ vs recessed contact length $L_{RC}$ for this device geometry.

### 3.5 Recessed Hall probes

We saw in Figure 5a that the use of recessed Hall probes in classical Hall bar devices eliminates the current perturbation effect. In this case the recessed contact length $L_{RC}$ is large compared to the recessed contact width $W_{RC}$. Furthermore, $W_{RC}$ is much smaller than the device dimensions $L$ and $W$. Some nanoscale devices such as nanocrosses [27,35] feature recessed contacts but do not always fulfil the conditions above. Here, $L_{RC}$ may be less than $W_{RC}$ and $W_{RC} \approx W$. In this case, current perturbation effects can occur as we observed above for metal contacts that overlap the semiconductor channel. Figures 6a and b show simulations of current density for two 5 μm × 1 μm Hall devices with 0.5 and 0.1 μm long and 1 μm wide recessed Hall probes with metal contacts (yellow) at the end. The current density is significantly reduced at the device centre for the $L_{RC} = 0.1$ μm device as current is diverted through the contacts. Figure 6c shows that this leads to a reduction in $V_H^{sim}$, as we saw for metal contacts that overlap the channel. The resistance of the current path through the contacts rises as $L_{RC}$ is increased and the current perturbation is reduced. For this particular geometry $V_H^{sim}/V_H^{real} \approx 0.98$ for $L_{RC} = 1$ μm. A more comprehensive discussion of recessed contacts as well as simulations of a range of device geometries is provided in the Supporting Information. Overall, recessed Hall probes give significantly more accurate Hall voltage measurements than metal probes with channel overlap. However, current perturbation effects should be considered in samples with $2L_{RC} < W_{RC}$ and where $W_{RC}$ is of the order of $W$ and $L$.

### 3.6 Metal probe geometry

Figure 7a shows simulations of $V_H^{sim}/V_H^{real}$ for 5 μm long samples as a function Hall-probe gap $G_C$ normalized to the sample width $W$ of 1 μm. We first consider contacts with limited invasiveness. The 300 nm wide probe pair with high contact resistance (red, $\sigma_{CR} = 10$ S/m) and the 10 nm wide probe pair (yellow) both start at $V_H^{sim}/V_H^{real} \approx 1$ for $G_C/W = 0$ and linearly decrease to $V_H^{sim}/V_H^{real} \approx 0.5$ at $G_C/W = 1$. This behaviour is approximated well by Equation 3, which is plotted as a grey line. Wider invasive contacts exhibit a different behaviour (green $W_C = 0.3$ μm, blue $W_C = 1$ μm). $V_H^{sim}/V_H^{real}$ is significantly smaller than 1 for $G_C/W = 1$, where the Hall probes only contact the side walls of the sample. This is due to the current perturbation effect discussed in Section 3.4. We will examine this more closely in Figure 7b.

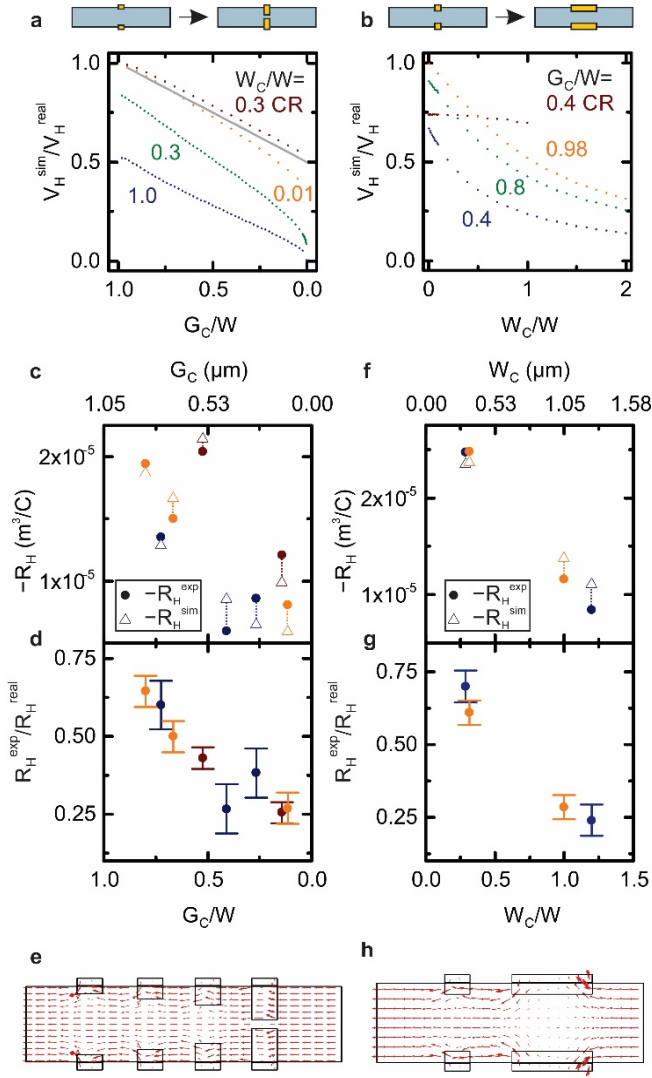

**Figure 7.** (a) $V_H^{sim}/V_H^{real}$ vs probe gap per sample width $G_C/W$ for 1 μm (blue), 0.3 μm (green), 10 nm (orange) wide probes with no contact resistance and 0.3 μm wide probes with high contact resistance (red). The grey line follows Equation 3. (b) $V_H^{sim}/V_H^{real}$ vs contact width per sample width $W_C/W$ for 0.4 μm (blue), 0.8 μm (green), 0.98 μm (orange) long contacts with no contact resistance, and 0.4 μm with high contact resistance (red). (c) Measured and simulated Hall coefficient $R_H^{exp}$ and $R_H^{sim}$ for different probe gaps $G_C$ from the device presented in Figure 2. The devices are represented in different colours. Experimental and corresponding simulated datapoints are connected by a dashed line for clarity. The carrier concentration for the simulations $n = 1.32 \times 10^{23}$ m⁻³ (red), $n = 2.78 \times 10^{23}$ m⁻³ (blue) and $n = 2.08 \times 10^{23}$ m⁻³ (orange) are based on the least-squares method for the best match of $R_H^{exp}$ and $R_H^{sim}$. (d) $R_H^{exp}/R_H^{real}$ vs $G_C/W$ where $R_H^{real}$ is based on the carrier concentrations obtained in (c). Error bars are based on the average difference between $R_H^{exp}$ and $R_H^{sim}$ in (c). (e) Simulation of current density for the device depicted in Figure 2a. (f) $R_H^{exp}$ and $R_H^{sim}$ for devices with different probe widths $W_C$. $R_H^{sim}$ is based on $n = 1.77 \times 10^{23}$ m⁻³ (blue) and $n = 1.54 \times 10^{23}$ m⁻³ (orange). (g) Corresponding $R_H^{exp}/R_H^{real}$. (h) Simulation of current density for the device shown in Figure 2c.

Figure 7b shows $V_H^{sim}/V_H^{real}$ as a function of probe width $W_C$. The Hall voltage should not depend on contact width in devices with non-invasive probes. This is consistent with the modelling of high contact resistance probes, where we obtain only a weak dependence on $W_C$ (red $G_C = 0.4$ μm). In stark contrast, $V_H^{sim}/V_H^{real}$ for invasive probes shows a strong dependence on $W_C$ even for very short Hall probe lengths (yellow $L_C = 10$ nm). This is because wider Hall probes draw more current, reducing the current density at the centre of the sample. For short probe pairs $V_H^{sim}$ decreases by approximately 50% at $W_C \approx W$, and even more for longer probe pairs (green $G_C = 0.8$ μm, blue $G_C = 0.4$ μm). This is particularly relevant for Hall measurements on nanowires where the channel width is often equal to or smaller than the probe width.[31,33,38,39]

Taking the insights gained through the simulations, we can now revisit the experimental data from the devices in Section 3.1. Figure 7c shows the experimentally obtained Hall coefficients $R_H^{exp}$ (circles) together with data from simulations $R_H^{sim}$ (triangles). The model geometry for the simulations is based on the dimensions of the real devices extracted from SEM and atomic-force microscopy (AFM). The model for the device from Figure 2a is shown in Figure 7e with a plot of the current density. Note that all probes are included in the model. This is important because the absolute position along the channel and position relative to other probe pairs can impact $V_H^{sim}$ (see Supporting Information S4). For our study it was important to use multiple Hall probe pairs on the same nanofin to isolate the effect of Hall probe geometry from sample-to-sample variations in carrier concentration. We found the best agreement between simulations and experiments using no contact resistance, confirming that the metal probes on InAs are strongly invasive. The carrier concentrations were adjusted in the simulations to match the amplitude of $R_H^{exp}$. We obtain $n = 1.32 \times 10^{23}$ m⁻³ (red), $n = 2.78 \times 10^{23}$ m⁻³ (blue) and $n = 2.08 \times 10^{23}$ m⁻³ (orange) using the least-squares method. The modelling fits the experimental data well. The average relative difference between the experimental and simulated datapoints is 16%. We attribute the discrepancy to two main factors. First, the shape of the metal contacts in the real device is slightly irregular with rounded edges and a slightly rugged outline (see Figure 2a). Second, the simulation assumes a perfectly uniform carrier concentration. In reality, surface accumulation layers[40] and variations in potential landscape, e.g., due to charge trapping[41,42] and polytypisms[43,44] likely lead to a more complex carrier distribution in InAs nanostructures. Overall, the data are in good agreement, validating the modelling. This allows us to estimate the ratio of the measured $R_H^{exp}$ and the real Hall coefficient $R_H^{real}$, with the latter analytically calculated using the carrier concentrations above. Figure 7d shows $R_H^{exp}/R_H^{real}$ versus $G_C/W$, with error bars based on the average difference between $R_H^{exp}$ and $R_H^{sim}$ in Figure 7c. We find that $R_H^{exp}$ is 60-70% of $R_H^{real}$ for large probe gaps and as low as 25% for the narrowest probe gaps. The latter would correspond to a four-fold overestimate of carrier concentration and underestimate of mobility. Specifically, for the device shown in Figure 2a, we estimate that the real mobility is approximately 2000 cm²/Vs. Without a correction we would obtain between 500 and 1250 cm²/Vs depending on which probe pair was used.

We obtain similar results for the two samples with probe pairs of different width $W_C$. Figure 7f shows the experimentally obtained data with simulations for $n = 1.77 \times 10^{23}$ m⁻³ (blue) and $n = 1.54 \times 10^{23}$ m⁻³ (orange). The corresponding estimates for $R_H{}^{exp}/R_H{}^{real}$ are shown in Figure 7g and give $R_H{}^{exp}/R_H{}^{real}$ as low as 0.25 for wide contacts. The effect of invasive contact width is well illustrated by the simulated current density in Figure 7h. The current density is significantly diminished between the wide contacts whereas it remains relatively unperturbed for the narrow contact pair.

## 4. Discussion and Conclusion

We have shown that compact sample geometries and invasive metal contacts significantly reduce the measured Hall voltage. This can lead to a substantial overestimate of the nanostructure's carrier density. For a typical $1 \times 1$ μm² nanofin with $300 \times 300$ nm² overlapping Hall probes, the measured Hall voltage is only ~20% of $V_H{}^{real}$ (G in Figure 3). We identified three contributions to the reduction in Hall voltage: (i) the rotation of the electric field vector to the Hall angle relative to the current density is not completed in low aspect ratio samples, leading to a reduction in Hall voltage; (ii) Contacts that draw no current but overlap the conduction channel will measure the average electrical potential of the probe-semiconductor interface, which always yields $V_H{}^{exp} \leq V_H{}^{real}$; (iii) Metal contacts with low to moderate contact resistance draw current and distort the current density in the nanostructure, which leads to further reduction in measured Hall voltage.

The contribution of (i) only depends on the sample aspect ratio $L{:}W$ and can therefore be estimated from the simulations provided in Figure 4. We recommend aiming for devices with an aspect ratio no smaller than 2:1 to keep the reduction of $V_H{}^{exp}$ to below 10%. Only one pair of Hall probes should be used at the centre of the device. Multiple Hall probe pairs are convenient for four-probe measurements, however, they often lead to Hall probes just fractions of $W$ away from the source and drain contacts.[19,20] This alone can lead to an underestimate of the Hall voltage by around 50%. The underestimate can be in excess of 80% when the invasiveness of the contacts is also taken into account. The contribution of (ii) only depends on the contact gap $G_C$ compared to the sample width $W$ and can be estimated using Equation 3. Estimating the contribution of (iii) is significantly more complicated because the current perturbation strongly depends on the contact width and length, but also on contact and sample resistance, and even sample thickness. Here, Equation 3 can be used as a lower bound in estimating a correction factor because the reduction in measured Hall voltage will always be larger for invasive contacts. We provide a table of $V_H{}^{sim}/V_H{}^{real}$ values for various sample geometries to help estimate the reduction in Hall voltage in the Supporting Information. Regardless, contact length and width should be decreased as much as possible to reduce the effect. Additionally, engineering probes with high contact resistance could reduce the current perturbation.

Overall, we find that Hall measurements even with invasive contacts are a suitable way to characterize 2D nanostructures. However, care should be taken to minimize and correct for the associated reductions in measured Hall voltage. While the considerations in this paper focussed on III-V nanostructures, we expect them to be relevant for measurements on other 2D-nanostructures such as graphene,[45] and transition metal dichalcogenites[10] with similar device geometries.

## Conflicts of interest

There are no conflicts to declare.

## Acknowledgements

This work was funded by the Australian Research Council (ARC) and the University of New South Wales. This work was performed in part using the NSW and ACT nodes of the Australian National Fabrication Facility (ANFF) and the Electron Microscope Unit (EMU) within the Mark Wainwright Analytical Centre (MWAC) at UNSW Sydney.

## Notes and References